\documentclass[aps,%preprint,
twocolumn,superscriptaddress]{revtex4}
\usepackage{amsfonts}
\usepackage{amsmath}
\usepackage{amssymb}
\usepackage{graphicx}
\usepackage{color}

\newcommand{\bq}{\begin{equation}}
\newcommand{\eq}{\end{equation}}

\begin{document}

\preprint{}
\title{Phenomenology of ESR in heavy fermion systems: Fermi liquid and
non-Fermi liquid regime }
\author{Peter W{\" o}lfle}
\affiliation{Institut f{\" u}r Theorie der Kondensierten Materie, Universit{\" a}t
Karlsruhe, D-76128 Karlsruhe, Germany}
\author{Elihu Abrahams}
\affiliation{Center for Materials Theory, Serin Physics Laboratory, Rutgers University,
Piscataway, NJ 08854-8019}
\date{\today }
\pacs{PACS number}

\begin{abstract}
We extend and apply a recent theory of the dynamical spin response of
Anderson lattice systems to interpret ESR data on YbRh$_{2}$Si$_{2}$.
Starting within a semiphenomenological Fermi liquid description at low
temperatures $T<T_{x}$ (a crossover temperature) and low magnetic fields $%
B\ll B_{x},$ we extend the description to the non-Fermi liquid regime by
adopting a quasiparticle picture with effective mass and spin susceptibility
varying logarithmically with energy/temperature, as observed in experiment.
We find a \textit{sharp} ESR resonance line slightly shifted from the local $%
f$-level resonance and broadened by quasiparticle scattering (taking unequal 
$g$-factors of conduction and $f$ electrons) and by spin-lattice relaxation,
both significantly reduced by the effect of ferromagnetic fluctuations. A
detailed comparison of our theory with the data shows excellent agreement in
the Fermi liquid regime. In the non-Fermi liquid regime we find a close
relation of the $T$-dependence of the specific heat/spin susceptibility with
the observed $T$-dependence of line shift and linewidth.
\end{abstract}

\maketitle

%\preprint{HEP/123-qed}

\section{Introduction}

In several recent experiments \cite{YRS1,Ce}, low-temperature ESR has been
observed in some heavy-fermion metals, in particular YRh$_{2}$Si$_{2}$ (YRS) 
\cite{YRS1,YRS2}. The phase diagram of YRS has a magnetic-field induced
quantum critical point and is a model system for the study of quantum
criticality in the Kondo lattice. Consequently, the observation of a narrow
ESR resonance in this compound aroused great interest, especially since it
was commonly believed that heavy-fermion ESR would be unobservable due to an
enormous intrinsic linewidth $\Delta B$ of order $k_{B}T_{K}/g\mu _{B}$ \cite%
{YRS1}. Here $T_{K}$ is the lattice coherence (``Kondo") temperature for the
onset of heavy-fermion behavior and $g\mu_{B}$ is the gyromagnetic ratio for
the resonance. These were the first observations of ESR in Kondo lattice
systems at $T<T_{K}$.

A common feature of the compounds in which ESR has been observed appears to
be the existence of ferromagnetic fluctuations \cite{Ce,FM} These findings
challenge our understanding of heavy fermion compounds: How does a sharp
electron spin resonance emerge despite Kondo screening and spin lattice
relaxation, and why is this process influenced by ferromagnetic
fluctuations? In a recent paper (\textquotedblleft AW") \cite{AW}, we
discussed the background of these questions and answered them in the
framework of Fermi-liquid theory. An alternative explanation based on
localized spins was subsequently proposed by Schlottmann \cite{schlottmann}.
The general derivation of Fermi liquid theory from the microscopic theory
for a two-band Anderson lattice model has been given by Yip \cite{Yip}

In YRS, the observed narrow dysonian \cite{feher} ESR line shape was
originally interpreted \cite{YRS1} as indicating that the resonance was due
to local spins at the Yb sites. Therefore, initially the authors speculated
that the narrow ESR line might indicate the suppression of the Kondo effect
near the quantum critical point, since, as explained above, carrying over
Kondo impurity physics to the Kondo lattice, one might expect the local
spins to be screened by the Kondo effect, giving rise only to a broad spin
excitation peak, too wide to be observed in ESR experiments. However, a
closer look \cite{ea} revealed that itinerant (heavy) electron ESR could
give rise to a similar line shape since the carrier diffusion in YRS is
quite slow. Thus, whether the resonance is that of localized or itinerant
spins remained an open question.

In this paper, we extend our previous work \cite{AW} to the non-Fermi liquid (NFL) region of the YRS phase diagram and make a detailed comparison with the data. Excellent agreement
is obtained for the Fermi-liquid (FL) regime. In
particular, the ratio of the contributions $\propto T^{2}$ and $\propto
B^{2} $ to the linewidth in the FL region is very well reproduced. In addition, we account for the anomalous behavior observed in the NFL region for the resonance line shift and the linewidth, One absolutely essential aspect of our theory is the lattice coherence of the quasiparticles in the Anderson or Kondo lattice model: it is this lattice coherence that is responsible for the absence in the lattice case of the strong local spin relaxation that is observed in single Kondo impurity physics. Attempts to account for the observed logarithmic temperature dependence of the lineshift as arising from single Kondo ion physics above the Kondo temperature are therefore problematic, since lattice coherence is lost in that case

\section{ESR in the Kondo-screened Anderson lattice model: Fermi liquid
regime.}

This was analyzed in Sec.\ III of AW. The Hamiltonian of the simplest
Anderson lattice model, assuming momentum independent hybridization is given
by 
\begin{eqnarray}
H &=& \sum_{\mathbf{k},\sigma }\epsilon _{\mathbf{k}\sigma }c_{\mathbf{k}\sigma
}^{+}c_{\mathbf{k}\sigma }+\sum_{i,\sigma }\epsilon _{f\sigma }n_{fi\sigma
}+U\sum_{i}n_{fi\uparrow }n_{fi\downarrow } \notag \\
&+& V\sum_{i,\mathbf{k},\sigma }(e^{i%
\mathbf{{k\cdot R}_{i}}}f_{i\sigma }^{+}c_{\mathbf{k}\sigma }+h.c.),
\end{eqnarray}%
where $\epsilon _{\mathbf{k}\sigma }=\epsilon _{\mathbf{k}}-\omega
_{c}\sigma /2,$ $\sigma =\pm 1,$ is the conduction-electron energy spectrum
and $\omega _{c}=g_{c}\mu _{B}B$ is its Zeeman splitting; $c_{\mathbf{k}%
\sigma }^{+},f_{i\sigma }^{+}$ are creation operators of the conduction
electrons in momentum and spin eigenstates $(\mathbf{k}\sigma )$, and of
electrons in the local $f$ level at site $\mathbf{{R}_{i}}$, respectively.
The operator $n_{fi\sigma }=f_{i\sigma }^{+}f_{i\sigma }$ counts the number
of electrons on the local level, and\ $\epsilon _{f\sigma }=\epsilon
_{f}-\omega _{f}\sigma /2$. $V$ and $U$ are the hybridization amplitude and
the Coulomb interaction matrix element. We take the Zeeman splittings $%
\omega _{c}$ and $\omega _{f}$ to be unequal, as they are in real materials.
We consider the limit $\omega _{f}-\omega _{c}\rightarrow 0$ in the Appendix.

We now review the results for the dynamical spin susceptibility $\chi
^{+-}(\Omega )$ obtained in AW \cite{AW}. We find a single resonance peak

\begin{equation}
\chi ^{+-}(\Omega )=\chi ^{+-}(0){\frac{-\omega _{r}+i\Gamma}{\Omega -\omega
_{r}+i\Gamma}},
\end{equation}%
where the resonance frequency is given by \cite{expl}

\begin{equation}
\omega _{r}=\omega _{f}-\frac{m}{m^{\ast }}(\omega _{f}-\omega _{c}),
\label{shift}
\end{equation}%
Here $m^{\ast }/m$ is the quasiparticle effective mass ratio. We note that
for equal $g$-factors the line position is not shifted. In the Appendix, we
discuss the complete result, showing that even in the case of unequal $g$%
-factors there is only a single resonance peak. We also show that the
residual Fermi liquid interaction effects drop out of the resonance
frequency.

The linewidth $\Gamma$ has contributions from quasiparticle scattering and
from the conduction electron spin lattice relaxation $\gamma $

\begin{equation}
\Gamma =A[\alpha (\pi T)^{2}+\frac{1}{4}(R\omega _{f})^{2}]\frac{1}{R}%
+2\gamma \frac{m}{m^{\ast }}\frac{1}{R}.  \label{width}
\end{equation}%
Here $R=[1+\widetilde{U}\chi ^{+-}(0)]$ is identified as the Wilson ratio, $%
\widetilde{U}$ is the Fermi-liquid spin-exchange interaction \cite{FL} and $%
\chi ^{+-}(0)=M/B$ is the static transverse spin susceptibility ($M$ is the
spin polarization). The numerical coefficient $\alpha $ depends on the band
structure and is of order unity. In the case of a sizeable ferromagnetic
interaction, ($\widetilde{U}>0$), $R>>1,$ the linewidth gets narrowed by a
factor $1/R.$ We suggest that this effect is responsible for the fact that
so far an ESR line has only been observed in compounds that exhibit
signatures of ferromagnetic fluctuations.

\subsection{Magnetic anisotropy of ESR line%\protect\bigskip
}

The magnetic response of YRS is strongly anisotropic, largely because of the
single ion anisotropy. The Zeeman Hamiltonian of the $f$-electron ground
state doublet of Yb in tetragonal symmetry has the form $H_{Z}=-\mu
_{B}g_{f\perp }(S_{x}B_{x}+S_{y}B_{y})-\mu _{B}g_{f\parallel }S_{z}B_{z}$ ,
where the $z$-axis is along the crystallographic $c$-axis. The anisotropy of
the $g$-factor is about a factor of $20$, $g_{f\perp }=3.6$ and $%
g_{f\parallel }=0.17$ . We assume for the present that the anisotropy of the
interaction is negligible. The Hamiltonian is then diagonal in the
coordinate system in the spin space that diagonalizes $H_{Z}$. The
eigenvalues of $H_{Z}$ are found as

\begin{equation}
\omega _{f}(\phi )=\mp \mu _{B}B\sqrt{g_{f\parallel }^{2}\cos ^{2}\phi
+g_{f\perp }^{2}\sin ^{2}\phi }\ ,  \label{anis}
\end{equation}%
where $\phi $ is the angle between the magnetic field ${%
\mathbf{B}}$ and the $c$-axis.

The above results, Eq.\ (2), for the dynamical spin susceptibility $\chi
^{+-}(\Omega )$ obtained in \cite{AW} for the isotropic model may then be
generalized to the anisotropic model by replacing $\omega _{f}$ by $\omega
_{f}(\phi )$ and taking the tensor of spin susceptibility projected onto the
direction of the static magnetic field. According to Ref.\ \cite{ganis}, the
angle dependence of the resonance frequency is well-represented by Eq.\ (\ref%
{anis}). This indicates that the anisotropic part of the residual
Fermi-liquid spin-exchange interaction is small. As we shall see later, the
temperature dependence of the line shift in the non-Fermi liquid regime
suggests that there may be a small anisotropic interaction component. We
shall explore the consequences of such a non-spin rotation invariant term in
Sec.\ IIC, below.

\subsection{ESR line shift and linewidth in the Fermi liquid regime}

In Fig.\ 1, we show a sketch of the phase diagram of YRS, including the $%
B,\; T$ ranges in which ESR experiments have been carried out. Here, the
crossover to the Fermi liquid (FL) regime is determined by the onset of FL
behavior in thermodynamic quantities \cite{see}. The $T^*$ crossover is
primarily determined by Hall effect results \cite{hall, sci} that can be
interpreted as a transition (from left to right) to a large Fermi surface.

\begin{figure}[ht]
\centerline{\includegraphics[width=0.8\linewidth, angle=90]{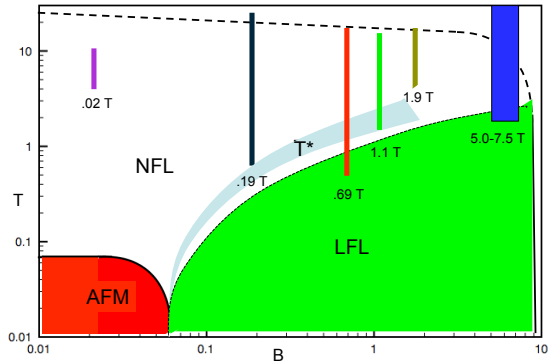}} %
\vskip -1.5 cm
\caption{Phase diagram of YRS showing field and temperature ranges of ESR
experiments}
\end{figure}

The high magnetic field ESR data reported in \cite{YRS2} show a crossover
from a low-temperature Fermi-liquid (FL) like regime to a higher temperature
non-Fermi liquid behavior at a temperature $T_{x}\simeq 5$ Kelvin. In the FL
regime the line shift appears to be temperature independent. Relative to the
ionic $g$-factor of $3.86$ \cite{YRS1}, the resonance is shifted to lower
values $g_{\perp}\simeq 3.42$ independent of magnetic field in the range $%
5.15$ to $7.45$ Tesla. Estimating the effective mass ratio, which is also
temperature independent below $T_{x}$\ in the magnetic field range
considered, as $m^{\ast }/m\sim 40$, we obtain from Eq.\ (\ref{shift}) a $g$%
-shift $\Delta g\simeq 0.04$, which is an order of magnitude too small. This
discrepancy may point to an additional small anisotropic spin interaction,
which we consider in the following subsection.

As for the linewidth, in the FL regime Schaufu\ss , et al \cite{YRS2} find a
linewidth that follows the law $\Gamma (T,B)\sim T^{2}$, extrapolating to $%
\Gamma (0,B)\sim B^{2}$ as $T\rightarrow 0$. The experimental ratio of the
prefactors of the $T^{2}$ and $B^{2}$ terms, $r_{exp}=B^{2}[\Gamma
(T,B)-\Gamma (0,B)]/[\Gamma (0,B)T^{2}]$ turns out to be $r_{exp}\sim 2$.
Estimating the Wilson ratio from the available specific heat data \cite%
{trov,NJP} at $T_{x}=5$ K and in magnetic fields $B\approx 6$ Tesla, $\Delta
C/T=0.032$ K$^{-1}$Yb$^{-1}$ and spin susceptibility data \cite{NJP,JPS} $%
\Delta \chi =M/B=0.224\mu _{B}^{2}$ K$^{-1}$Yb$^{-1}$ as $R=[\chi /(g_{f}\mu
_{B}/2)^{2}](\pi ^{2}T/3\Delta C)\simeq 7.5$, we calculate from Eq.\ (\ref%
{width}) the theoretical ratio $r_{th}\sim 1.2\alpha $, in good agreement
with the experimental value. Note that the large enhancement of the single
particle Zeeman splitting by the Fermi liquid interaction (a factor $R$) is
essential in obtaining this agreement.

\subsection{Effect of non spin-rotation invariant Fermi liquid interaction.}

The spin-orbit interaction in conjunction with the tetragonal lattice
anisotropy may be expected to lead to a small admixture of a
non-spin-symmetric component to the Fermi liquid interaction of the form $%
-4I(\overrightarrow{S}\cdot \hat{c})^{2}$ , where $\hat{c}$ is the unit
vector along the $c$-axis of the tetragonal lattice. Taking the magnetic
field along the $b$-axis, we employ a coordinate system in spin space, in
which the $z$-axis is oriented along the magnetic field and the $x$-axis
along the $c$-axis (see Fig.\ 2). %\vskip-1.2cm 
\begin{figure}[th]
\centerline{\includegraphics[width=0.98\linewidth,angle=90
]{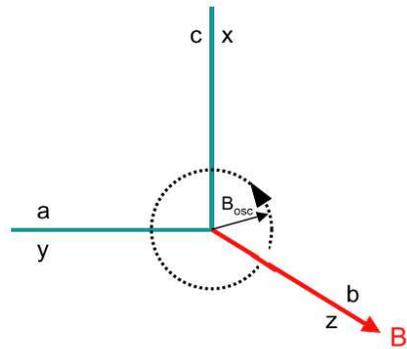}} \vskip-3cm
\caption{Specification of axes for magnetic fields in ESR experiment}
\end{figure}

The ESR oscillating transverse magnetic field is circularly polarized in the 
$x-y$ plane, which is the $a-c$ plane of the crystal,
perpendicular to the static magnetic field. The screening of the static
magnetic field is effected in linear order in $I$ only when the component of
the magnetic field along $\hat{c}$ is nonvanishing. In turn, the dynamic
screening is changed at linear order in $I$, for any component of $\mathbf{B}%
_{static}$ perpendicular to $\hat{c} $ . Thus, the static and dynamic
screening are effected differently by $I$, which gives rise to a resonance
line shift, which we now calculate. As we show in the Appendix, the
dynamical screening of the $ff$ component of the dynamical susceptibility is
modified in the presence of $I$ to 
\begin{equation*}
\chi _{ff}^{+-}(i\Omega _{m})=\chi _{ff,H}^{+-}(i\Omega _{m})[1+\widetilde{U}%
_{d}\chi _{ff}^{+-}(i\Omega _{m})]
\end{equation*}%
Thus, in the transverse spin response, the Fermi-liquid interaction is
changed to $\widetilde{U}_{d}=\widetilde{U}+I\sin ^{2}\phi $. The notation
is as in AW \cite{AW}, Eqs.\ (16-18): $\chi _{ff,H}$ is the susceptibility
of Fermi-liquid quasiparticles in the absence of vertex corrections (bubble
diagram only) and the subscript $H$ indicates that the bare Zeeman energy $%
\omega _{f}$ is replaced everywhere by 
\begin{equation*}
\widetilde{\omega }_{f}=\omega _{f}[1+\widetilde{U}_{s}\chi
_{ff}^{+-}(0)]=\omega _{f}[1-\widetilde{U}_{s}\chi _{ff,H}^{+-}(0)]^{-1}.
\end{equation*}%
Here ${\widetilde{U}}_{s}= {\widetilde{U}}+4I\cos^2\phi$ and ${\widetilde{U}}
$ is the renormalized onsite $ff$ repulsion that appears in the effective
low-energy Hamiltonian (see AW for further details). The resonance position
is therefore shifted as

\begin{equation*}
\omega _{r}=\widetilde{\omega }_{\mathbf{k}_{F}}^{-}[1-\widetilde{U}_{d}\chi
_{ff,H}^{+-}(0)]=\omega _{k_{F}}^{-}\frac{1-\widetilde{U}_{d}\chi
_{ff,H}^{+-}(0)}{1-\ \widetilde{U}_{s}\chi _{ff,H}^{+-}(0)}
\end{equation*}%
Substituting $\omega _{k_{F}}^{-}$ as obtained in the Appendix and using the
definition of $R$, we find that the resonance frequency is modified from
Eq.\ (3) to 
\begin{eqnarray}
\omega _{r}&\simeq&\omega _{k_{F}}^{-}[1-I\chi _{ff,H}^{+-}(0)R] \notag \\
&\simeq&\omega _{f}-\frac{m}{m^{\ast }}(\omega _{f}-\omega _{c})-\omega
_{f}I(1-5\cos ^{2}\phi )\chi _{ff}^{+-}(0), \notag
\end{eqnarray}
or
\begin{equation}
g_{r} \simeq g_{f}-\frac{m}{m^{\ast }}%
(g_{f}-g_{c})-g_{f}I(1-5\cos ^{2}\phi )\chi _{ff}^{+-}(0).
\end{equation}

Since most of the ESR data have been taken in the configuration of magnetic
field perpendicular to the $c$-axis, \textit{i.e}. $\phi =\pi /2$ ,
we concentrate on this case from now on. We may try to determine $I$ by
fitting the low-temperature line shift \cite{YRS2}. In the following, we
assume $I$ to be independent of temperature and magnetic field. In the NFL
regime, experimentally $\chi _{ff}^{+-}(0;T)$ is a decreasing function of
temperature, such that the $g$-shift increases as observed in experiment,
provided $I>0$. We may relate $I$ to $\chi _{ff}^{+-}(0;T_{1})$ at a
reference point $\ T_{1}=4$K, $B=0.2$T, where $\chi
_{ff}^{+-}(0;T_{1})\approx 1.3\times 10^{-6}$m$^{3}$/mol \cite{trov2} as $%
I=(g_{f}^{ion}-g_{r})/(g_{f}\chi _{ff}^{+-})$ [the reference $g$-factor $%
g_{f}^{ion}$ is actually reduced by the factor $(1-m/m^{\ast }$)]. The data
at low magnetic fields, $B=0.18$T and $B=0.68$T show a $g$-factor of $%
g\simeq 3.5$ at the lowest temperature, $T=2$K, whereas the high-field data
show $g=3.42$ in the Fermi liquid regime. It follows that $I\approx
0.075\times 10^{6}$ m$^{-3}$mol.\ \ From a comparison with the temperature
dependence of the resonance frequency we determine in the next Section a
value of $I=0.063\times 10^{6}$ m$^{-3}$mol, which agrees very well with the
independently-obtained value above. We observe in passing that the magnetic
susceptibility data indicate that in the non-Fermi liquid regime $\widetilde{%
U}$ appears to depend on both temperature and magnetic field.

\section{ESR in the non-Fermi liquid regime.}

We now attempt to phenomenologically relate the framework we have set up to
the ESR data in the NFL regime by using the observed specific heat and
susceptibility $T$ and $B$ dependences. The non-Fermi liquid behavior in the
temperature range $T>T_{x}$ appears in the ESR data as a nearly logarithmic
increase of the $g$-factor with temperature and a change in the temperature
dependence of the linewidth from $T^{2}$ to $T$ . This change into the NFL
regime occurs at about the same temperature as the observed changes in the
specific heat and spin susceptibility.

\subsection{Resonance shift}

In the theoretical resonance shift, Eq.\ (6), the $T$- and $B$- dependences
enter in two ways, if we continue to assume that the temperature dependence
of the anisotropic Fermi-liquid interaction parameter $I$ may be neglected:
1) through the susceptibility $\chi _{ff}^{+-}(0)$, which we get from
experiment, and 2) through the effective mass ratio, which we extract from
the measured specific heat $\gamma $-coefficient ($\Delta C=\gamma T$), by
taking $\gamma \propto m^{\ast }/m$. Using these experimentally determined
quantities, we shall use Eq.\ (6) to evaluate the theoretical resonance
shift and compare it to the observed one. 
%As discussed above, the lineshift caused by the first term $\propto m/m^{\ast }$ is %small. We shall concentrate on the last term in Eq.\ (6).

We shall use Eq.\ (6) to calculate the $g$-shift $\delta g=g-g_{f}^{ion}$ at
two reference temperatures $T=4$K and $T=10$K. The inputs are the values of $%
m/m^{\ast }$ from the observed specific heat \cite{trov}, taking $m^{\ast
}/m=40$ at $T=5$ K \ and $B=6$ T as a reference point, the observed
susceptibility \cite{JPS, trov2} and the value of the anisotropic FL
interaction $I$. These data and the calculated $g$-shifts are collected in
Table I. Since we have assumed that $I$ is independent of $T$ and $B$, we
can evaluate it from Eq.\ (6) using experimental data at $T=4$K, $B=0.2$T as
discussed in the previous Section. The result is $I=0.063\times 10^{6}$ mol/m%
$^{3}$ 
%We may fit the spin susceptibility data from Fig.\ 6 of \cite{JPS}\ and
%Fig.\ 1 of \cite{trov2} to the interpolation formula $\chi =M/B=[a-b\ln
%T]/[1+c\ln T]$ 10$^{-6}$m$^{3}$/mol. The $B$-dependent coefficients are
%given by $a=1.96-0.42\ln B-0.063\ln ^{2}B,$ \ $b=0.52-0.085\ln B-0.032\ln
%^{2}B$ \ $c=-0.125+0.0081\ln ^{2}B$. Here $T$ is in Kelvin and $B$ is in Tesla.
%With the interpolation formula, we determine $\chi $ at the values of temperature and %magnetic field needed to calculate the line shift given by Eq.\ (6).
The data on the ESR line shift given in Fig.\ 2 of \cite{YRS2} show an
approximately linear $\ln T$ dependence in the $T$-range $4K<T<10K$.
Therefore, to check the accuracy of our theoretical result, Eq.\ (6), we fit
the two calculated $\delta g$ values to a linear $\ln T$ function and give
the resulting theoretical slope in the last column of Table I. The
comparison of the calculated and observed values of the slope $\Delta
g/\Delta \ln T$ is shown in Fig.\ 3. It is seen that the agreement is quite
good, supporting our assumption of a constant interaction $I$. The theory
explains the rather strong dependence of the $\ln T$ term on magnetic field,
decreasing by approximately a factor of $8$ as the magnetic field is stepped
up from $0.19$ T to $7.45$ T. The present theory predicts that the slope of
the $\ln T$ dependence of the $g$-shift depends sensitively on the magnetic
field orientation (the angle $\phi $) and it reverses sign when $\mathbf{B}%
_{static}$ is oriented along the ${\hat{c}}$-axis of the crystal. 
\begin{table}[h]
\caption{ Experimental values of susceptibility and specific heat
coefficient and calculated $g$-shift [from Eq.\ (6)] at different $B$ and
two temperatures $T_{1}=4$K, $T_{2}=10$K. We chose $I=0.063\times 10^{6}$
mol/m$^{3}$. Units: $\protect\chi $ in $10^{-6}$m$^{3}/$mol, $\protect\gamma 
$ in J/(mol K$^{2}$) }
\begin{center}
\begin{ruledtabular}
\begin{tabular}{c c c c c c p{1.5 cm} | l}
$B$(T) &$\chi(T_1)$&$\chi(T_2)$ & $\gamma(T_1)$ & $\gamma(T_2)$ & $\delta g (T_1)$ & $\delta g (T_2)$ & $(\Delta g/\Delta\ln T)_{th}$ \\
\hline 
7.5  & 0.82 & 0.65 &0.24& 0.185 & -0.222 &-0.197& 0.027  \\
6.0  & 0.91 & 0.70 &0.26& 0.18 & -0.239 &-0.210& 0.032  \\
5.0  & 0.98 & 0.74 &0.27& 0.17 & -0.254 &-0.222& 0.035 \\
1.85  & 1.13 & 0.78 &0.29& 0.15 & -0.283 &-0.238& 0.050 \\
1.0  & 1.2 & 0.80 &0.29& 0.15 & -0.298 &-0.243& 0.060\\
0.68  & 1.23 & 0.80 &0.29& 0.15 & -0.305 &-0.243& 0.068 \\
0.5  & 1.25 & 0.80 &0.29& 0.15 & -0.309 &-0.243& 0.072 \\
0.2  & 1.3 & 0.80 &0.29& 0.15 & -0.320 &-0.243& 0.085 
\end{tabular}
\end{ruledtabular}
\end{center}
\end{table}

\begin{figure}[h]
\centerline{\includegraphics[width=0.9\linewidth, angle=90]{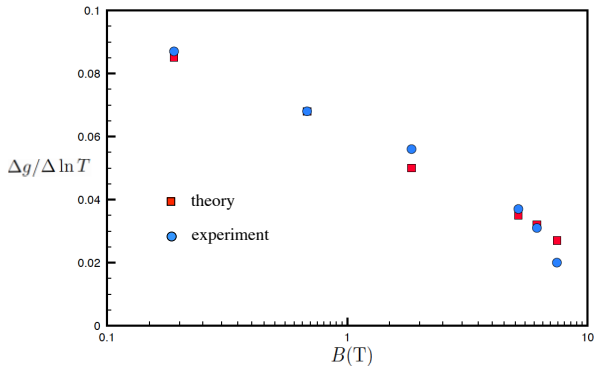}} %
\vskip -1.4 cm
\caption{Comparison of $g$-shift slopes for different $B$. The theoretical
and experimental values are identical at $B=0.68$T}
\end{figure}

\subsection{Linewidth}

Turning now to the linewidth, we use the analyticity properties of the self
energy $\Sigma (\omega )$ to infer the linewidth from the temperature
dependence of the specific-heat coefficient $\gamma (T)$. Here one has to
observe that only part of the specific heat enhancement is coming from the
nonanalytic contribution of the self energy $\Sigma $. An additional part is
coming from the regular (analytic) contribution to $\Sigma $. Therefore one
may split the effective mass into two components, $m^{\ast }/m=(m^{\ast
}/m)_{reg}+(m^{\ast }/m)_{sing}$. The specific heat data show a $\ln T$
variation over a wide range on top of a background. If we identify the
background with $(m^{\ast }/m)_{reg}$, the singular part at the reference
point $T=5$ K and $B=6$ T is about $60\%$ of the total, i.e. $(m^{\ast
}/m)_{s}=24$, taking the FL value $m^{\ast }/m=40$. The singular part can
now be associated with the non-Fermi-liquid logarithmic temperature
dependence of $\gamma $. Thus, $\gamma _{sing}\propto (m^{\ast }/m)_{sing}$.
To account for the crossover from NFL to FL behavior at $T=T_{x}$, we adopt
an interpolation formula $\gamma _{sing}=-c\ln [(T^{2}+T_{x}^{2})/T_{0}^{2}]$%
. Since $(m^{\ast }/m)=[1-\mathrm{Re}\{\partial \Sigma /\partial \omega
\}|_{0}]$, we may write $(m^{\ast }/m)_{sing}=-\mathrm{Re}\{\partial \Sigma
_{sing}/\partial \omega \}|_{0}$. The temperature dependence $(m^{\ast
}/m)_{sing}=-a\ln [(T^{2}+T_{x}^{2})/T_{0}^{2}]$ may be approximately
converted into a frequency dependence $\mathrm{Re}\{\partial \Sigma
_{sing}/\partial \omega \}=a\ln [(\omega ^{2}+T_{x}^{2})/T_{0}^{2}]$ of the
nonanalytic real part of the self energy. The self energy in the complex
plane may be inferred as

\begin{equation}
\Sigma _{sing}(\omega )=2a\omega \ln [(-i\omega +T_{x})/T_{0}]  \notag
\end{equation}%
From this approximate model the imaginary part of the self energy and hence
the quasiparticle contribution [the first term in Eq.\ (4)] to the ESR line
width follows as

\begin{equation}
\Gamma _{qp}=2\frac{m}{m^{\ast }}\frac{1}{R}\mathrm{Im}\Sigma _{s}(\omega
=T)=pT\tan ^{-1}(T/T_{x})  \notag
\end{equation}%
In the limit of $T\ll T_{x}$ the above expression recovers the Fermi liquid
result $\Gamma _{qp}\propto T^{2}$, as discussed above, while at $T\gg T_{x}$
the non-Fermi liquid result $\Gamma _{qp}\propto T$ is obtained. This is in
qualitative agreement with experiment. It is worth pointing out that a
similar structure of the self energy has been proposed for the
\textquotedblleft strange metal" phase of the cuprates under the name
\textquotedblleft marginal Fermi liquid theory" \cite{mfl}. By comparison
with the effective mass ratio we find $a_{th}\approx 6$ and using $R\approx 7
$ we get $p_{th}\approx 0.1$. The experimental value is $p_{ex1}\approx 0.02$%
, which is quite a bit smaller. Similarly, from the experimentally observed
coefficient of the $T^{2}$ term of the linewidth, $\Gamma /T^{2}\approx 0.004
$ K$^{-1}$ in the Fermi liquid regime, one extracts again a value $%
p_{ex2}\approx 0.02$ . The discrepancy may come from our assumption that $%
\Sigma _{sing}$ is entirely due to spin-flip scattering and from our very
approximate determination of $\Sigma _{sing}$. Vertex corrections will for
example remove any non-spin interaction contribution to $\Sigma _{sing}$
from the linewidth $\Gamma $. If this is correct, it would imply that the
fluctuations contributing most to the $\ln T$ term in the specific heat are
nonmagnetic in origin. Finally we comment on the possible contribution to $%
\Gamma $ caused by the regular part of $\Sigma _{reg}(\omega )$. In this
case the prefactors $c_{r},c_{i}$ of the low energy limiting forms $\mathrm{%
Re}\Sigma _{reg}(\omega )=\Sigma _{reg}(0)+c_{r}\omega $ and $\mathrm{Im}%
\Sigma _{reg}(\omega )=c_{i}\omega ^{2}$ are not directly related. The
Kramers-Kronig relations imply in this case that, e.g., the higher frequency
parts of $\mathrm{Re}\Sigma _{reg}(\omega )$ will predominantly determine
the coefficient $c_{i}$, while the coefficient $c_{r}$ has little influence
in this. In the present case, the resulting imaginary part and coefficient $%
c_{i}$ is apparently small.

\section{Conclusion}

We extended and applied our recent theory \cite{AW} of the dynamical spin
response of Anderson lattice systems to interpret ESR data on YbRh$_{2}$Si$%
_{2}$. Starting with a semiphenomenological Fermi-liquid description at low
temperatures $T<T_{x}$ (a crossover temperature) and low magnetic fields $%
B\ll B_{x},$ we extended the description to the non-Fermi liquid regime by
adopting a quasiparticle picture with effective mass and spin susceptibility
varying logarithmically with energy/temperature, as observed in experiments.
We find a \textit{sharp} ESR resonance line that is broadened by
quasiparticle scattering and spin lattice relaxation, both significantly
reduced by the effect of ferromagnetic fluctuations. A more complete
evaluation of the results presented in our first paper shows that the
ESR-line position is shifted by an amount $\propto (g_{f}-g_{c})$ , thus
reducing to zero for equal $g$-factors. In the case of different $g$-factors
there is only one sharp resonance line at $\omega \simeq \omega _{f}$, the
local $f$-resonance frequency. The observed strong anisotropy of the ESR
response is shown to follow from the single-ion spin anisotropy, assuming an
approximately spin-conserving exchange interaction.

A detailed comparison of our theory with the data shows excellent agreement
in the Fermi-liquid regime, when the model is amended by a small anisotropic
part of the spin exchange interaction, induced by spin-orbit coupling. We
assumed the strength of the latter to be independent of temperature and
magnetic field throughout the regime considered in the experiments. In
particular, the ratio of the contributions $\propto T^{2}$ and $\propto
B^{2} $ to the linewidth in the FL region is very well reproduced by theory.

In the non-Fermi liquid regime we find a close relation of the $T$%
-dependences of the specific heat and spin susceptibility with the observed $%
T$-dependences of the line shift and linewidth. There are two terms
contributing to the temperature dependence of the lineshift [Eq.\ (6)] in
opposite ways. The first and dominant one is proportional to the spin
susceptibility $\chi $ and leads to a resonance frequency increasing with
temperature, while the second and smaller one is proportional to the inverse
specific heat coefficient $1/\gamma $, leading to a decreasing behavior. The
observed approximately linear $\ln T$ dependence of the $g$-shift is
determined by $\chi $ and $1/\gamma $, (in the restricted temperature
regime, where the remaining curvature in both $\chi $ and $1/\gamma $ tends
to compensate), while the magnitude of the shift is fitted by adjusting the
anisotropic exchange interaction constant $I$. The observed rather strong
magnetic field dependence of the prefactor of $\ln T$ is very well accounted
for by $\chi $. Finally, we attempted to relate the linewidth to the
singular part of the self energy $\Sigma _{sing}$, by identifying the $\ln T$
contribution to the specfic heat coefficient with the effective mass deduced
from $\Sigma _{sing}$. On a qualitative level, the observed crossover from $%
T^{2}$ to linear $T$ behavior of the linewidth upon entering the non-Fermi
liquid regime is reproduced. However, the line width is found to be
approximately a factor of $5$ too large compared to experiment. The most
likely explanation of this discrepancy is our neglect of vertex corrections,
which would remove any non-magnetic contribution to $\Sigma _{sing}$ from
the linewidth.

Overall the extended quasiparticle picture used to account for the
ESR-properties in the non-Fermi liquid regime appears to work quite well. It
would be interesting to compare with data taken at much lower temperatures,
when the present theory would predict, e.g. nonlinear variation of the $g$%
-shift with $\ln T$, as exhibited by $\chi $. Also, a cleaner identification
of the linear $T$ dependence of the linewidth would be essential to
corroborate the extended quasiparticle picture. Finally, our theory makes
definite predictions for the anisotropy of the lineshift.

\begin{acknowledgements}
We thank the Aspen Center for Physics, where part of this work
was completed.  P.W.\ acknowledges partial support from the
DFG-Center for Functional Nanostructures and the DFG-Forschergruppe
``Quantum Phase Transitions".

\end{acknowledgements}

\appendix*
\section{}
%\end{verbatim}

\subsection{Derivation of dynamical susceptibility: spin-rotation invariant
Fermi liquid interaction}

In this Appendix we derive the quasiparticle properties and the dynamical
spin susceptibility in a more detailed and complete way than was done in AW 
\cite{AW}. We start from the Dyson equation for the single particle Green's
functions, 
\begin{widetext}
\begin{equation}
\mathcal{G}^{-1}\mathcal{G} = 
\begin{pmatrix}
i\omega _{n}-\epsilon _{f\sigma }-\Sigma _{f\sigma }(i\omega _{n},\mathbf{k})
& -V \\ 
-V & i\omega _{n}-\epsilon _{\mathbf{k\sigma }}-\Sigma _{c\sigma }(i\omega_n
,\mathbf{k})%
\end{pmatrix}%
\begin{pmatrix}
G_{\mathbf{k}\sigma }^{ff} & G_{\mathbf{k}\sigma }^{cf} \\ 
G_{\mathbf{k}\sigma }^{fc} & G_{\mathbf{k}\sigma }^{cc}%
\end{pmatrix}%
=\mathbf{1}.
\end{equation}
\end{widetext}
We approximate the conduction electron retarded self-energy by $\Sigma
_{c\sigma }(\omega +i0,\mathbf{k})=-i\gamma $, where $\gamma$ is the
conduction-electron spin-lattice relaxation rate. To carefully derive the
quasiparticle Zeeman energies, we make use of the conservation of total spin
in the model considered here, to remove the $f$-electron Zeeman term of the
Hamiltonian by performing a gauge transformation that shifts the zero of
energy of $\uparrow $-spins and $\downarrow $-spins by $\mp \omega _{f}/2$
respectively. As a consequence $i\omega _{n}$ is replaced by $i\omega
_{n}+\sigma \omega _{f}/2$ , $\epsilon _{f\sigma }$ by $\epsilon _{f}$ and
the conduction electron Zeeman energy is changed to $-\sigma (\omega
_{f}-\omega _{c})/2$. For an isotropic band structure ($\epsilon _{\mathbf{k}%
}=\epsilon _{k}$), the magnetic field dependence of the self energy
(neglecting small band edge terms) is then of the form 
\begin{equation*}
\Sigma _{f\sigma }(\omega +i0,\mathbf{k})\simeq \Sigma _{f}(\omega +\sigma
\omega _{f}/2+i0,\epsilon _{k}-\sigma (\omega _{f}-\omega _{c})/2).
\end{equation*}
In $\mathcal{G}^{-1}_{11}$, we expand this $f$-self energy about the Fermi
energy: 
\begin{widetext}
\begin{eqnarray}
\omega +\sigma \omega _{f}/2-\epsilon _{f}-\Sigma _{f\sigma }(\omega +i0,%
\mathbf{k}) &=&(\omega +\sigma \omega _{f}/2)(1-\partial \Sigma
_{f}/\partial \omega |_{0})-\epsilon _{f}-\Sigma _{f}(i0,\epsilon _{k_{F}}) 
\notag \\
&-&(\partial \Sigma _{f}/\partial \epsilon _{k}|_{k_{F}})(\epsilon
_{k}-\epsilon _{k_{F}}-\sigma (\omega _{f}-\omega _{c})/2)+i\mathrm{Im}%
\Sigma _{f}(\omega +i0,\epsilon _{k})  \notag \\
&=&z^{-1}[\omega -\widetilde{\epsilon }_{fk\sigma }+i\gamma _{fk}],
\end{eqnarray}
where $z^{-1}=1-\partial \Sigma_f /\partial \omega |_{0}$, $\widetilde{%
\epsilon }_{fk\sigma }=\widetilde{\epsilon }_{fk}-\sigma \omega
_{f}/2+\sigma (\partial \Sigma _{f}/\partial \epsilon _{k}|_{k_{F}})(\omega
_{f}-\omega _{c})/2$, $\widetilde{\epsilon }_{fk}=z[\epsilon _{f}+\Sigma
_{f}(i0,\epsilon _{k_{F}})+(\partial \Sigma _{f}/\partial \epsilon
_{k}|_{k_{F}})(\epsilon _{k}-\epsilon _{k_{F}})]$, and $\gamma _{fk}=z%
\mathrm{Im}\Sigma _{f}(\omega +i0,\epsilon _{k})$

Then for low energies one has a quasiparticle description, with $G_{\mathbf{k%
}\sigma }^{ff}(\omega )=z_{\sigma }\widetilde{G}_{\mathbf{k}\sigma }^{ff}$,
\ $G_{\mathbf{k}\sigma }^{cf}=\sqrt{z_{\sigma }}\widetilde{G}_{\mathbf{k}%
\sigma }^{cf}$ and \ the renormalized \ hybridization amplitude $\widetilde{V%
}^{2}$\ $=z_{\sigma }V^{2}$. The complex energy eigenvalues\ are given by 
\begin{equation}
\zeta _{\mathbf{k\sigma }}^{\pm }=\frac{1}{2}(\widetilde{\epsilon }_{f\sigma
}-i\gamma _{fk}+\epsilon _{\mathbf{k\sigma }}-i\gamma )\pm \sqrt{\frac{1}{4}(%
\widetilde{\epsilon }_{f\sigma }-i\gamma _{fk}-\epsilon _{\mathbf{k\sigma }%
}+i\gamma )^{2}+\widetilde{V}^{2}}=\epsilon _{\mathbf{k}}^{\pm }-\frac{1}{2}%
\sigma \omega _{\mathbf{k}}^{\pm }-i\gamma _{\mathbf{k}}^{\pm },
\end{equation}%
where, assuming $|\epsilon _{\mathbf{k}_{F}}|>>$ $|\widetilde{\epsilon }%
_{fk}|,\widetilde{V}$, 
\begin{eqnarray}
\epsilon _{\mathbf{k}}^{\pm }&=&\epsilon _{k_{F}}^{\pm }+\frac{1}{2}%
(\epsilon _{k}-\epsilon _{k_{F}})(1+z(\partial \Sigma _{f}/\partial \epsilon
_{k}|_{k_{F}}))[1\pm {\frac{\epsilon _{\mathbf{k}_{F}}-\widetilde{\epsilon }%
_{f}}{\sqrt{(\widetilde{\epsilon }_{f}-\epsilon _{\mathbf{k}_{F}})^{2}+4%
\widetilde{V}^{2}}}]},\cr \omega _{\mathbf{k}}^{\pm }&=&\omega _{f}+\frac{1}{%
2}[(\omega _{c}-\omega _{f})(1+z(\partial \Sigma _{f}/\partial \epsilon
_{k}|_{k_{F}}))[1\pm {\frac{\epsilon _{\mathbf{k}_{F}}-\widetilde{\epsilon }%
_{f}}{\sqrt{(\widetilde{\epsilon }_{f}-\epsilon _{\mathbf{k}_{F}})^{2}+4%
\widetilde{V}^{2}}}]}\cr \gamma _{\mathbf{k\sigma }}^{\pm }&=&\frac{1}{2}%
(\gamma _{fk}+\gamma )\mp \frac{1}{2}(\gamma _{fk}-\gamma )\frac{\epsilon _{%
\mathbf{k\sigma }}-\widetilde{\epsilon }_{f\sigma }}{\sqrt{(\widetilde{%
\epsilon }_{f}-\epsilon _{\mathbf{k}_{F}})^{2}+4\widetilde{V}^{2}}}
\end{eqnarray}
\end{widetext}

From now on we can safely neglect the term involving $(\partial \Sigma
_{f}/\partial \epsilon _{k}|_{k_{F}})$, since it is small - of order $z<<1$.

Using partial fraction decomposition, we construct the retarded Green
function 
\begin{equation}
\widetilde{G}_{\mathbf{k}\sigma }^{ff}(\omega +i0)={\frac{a_{\mathbf{k}%
\sigma }^{ff,+}}{\omega -\zeta _{\mathbf{k\sigma }}^{+}}}+{\frac{a_{\mathbf{k%
}\sigma }^{ff,-}}{\omega -\zeta _{\mathbf{k\sigma }}^{-}}}
\end{equation}%
and similar expressions for $\widetilde{G}_{\mathbf{k}\sigma }^{cf}$ and $G_{%
\mathbf{k}\sigma }^{cc}$, where, with $u_{\mathbf{k}\sigma }=\zeta _{\mathbf{%
k}\sigma }^{+}-\zeta _{\mathbf{k}\sigma }^{-}$, 
\begin{eqnarray*}
a_{\mathbf{k}\sigma }^{ff,\pm }&=&\pm (\zeta _{\mathbf{k\sigma }}^{\pm }-%
\widetilde{\epsilon }_{\mathbf{k}\sigma })/u_{\mathbf{k}\sigma }, \\
 a_{\mathbf{k}\sigma }^{cc,\pm }&=&\pm (\zeta _{\mathbf{k\sigma }}^{\pm }-\epsilon
_{f\mathbf{\sigma }})/u_{\mathbf{k}\sigma }\\
 a_{\mathbf{k}\sigma
}^{cf,\pm }&=&\pm \widetilde{V}/u_{\mathbf{k}\sigma }.
\end{eqnarray*}%
For sufficiently small imaginary parts, $\gamma \ll (\widetilde{V},%
\widetilde{\epsilon }_{f\sigma })$, we may neglect them in the weight
factors $a_{\mathbf{k}\sigma }^{ff,\pm },...$ and replace $\zeta _{\mathbf{%
k\sigma }}^{\pm }$ by $\epsilon _{\mathbf{k\sigma }}^{\pm }$.

The quasiparticles interact via the residual Fermi liquid interaction. For
ESR, the relevant component of the Fermi liquid interaction is the spatially
isotropic spin-antisymmetric part described by the Landau parameter $%
F_{0}^{a}=-2N_{0}\widetilde{U}$ . Here $\widetilde{U}$ \ is the coupling
constant of a spin isotropic exchange interaction $H_{ex}=-$\ $\widetilde{U}%
\overrightarrow{S}\cdot \overrightarrow{S}$ , which leads to a quasiparticle
energy shift $\delta \omega _{\mathbf{k}}=$\ $\widetilde{U}M$, where the
spin polarization $M$ is given by $M=\chi _{H}\widetilde{\omega }_{k_{F}}$,
with $\chi _{H}$ the unscreened static spin susceptibility (in the absence
of the Fermi liquid interaction) and $\widetilde{\omega }_{k_{F}}$ the fully
renormalized quasiparticle Zeeman splitting at the Fermi energy.\ 

For definiteness in the following we assume\ a band filling of somewhat less
than two electrons per site, such that the Fermi level lies in the lower
quasiparticle band (energy $\zeta _{\mathbf{k\sigma }}^{\pm }$), and $%
\widetilde{\epsilon }_{f\sigma }$ is close to the Fermi energy. The screened
Zeeman splitting is then obtained by solving the self-consistent equation $%
\widetilde{\omega }_{k_{F}}=\omega _{k_{F}}^{-}+$\ $\widetilde{U}\chi _{H}%
\widetilde{\omega }_{k_{F}}$. The solution is 
\begin{equation}
\widetilde{\omega }_{k_{F}}=\frac{\omega _{k_{F}}^{-}}{1-\ \widetilde{U}\chi
_{H}}=\omega _{k_{F}}^{-}R,
\end{equation}
where $R=1/(1-\ \widetilde{U}\chi _{H})$ is the Wilson ratio. One observes
that in the case of ferromagnetic correlations, when $R>>1$, the single
particle Zeeman splitting is enhanced by a large factor. We conjecture that
this effect should be observable in tunneling experiments. In the
two-particle spectrum probed by electron spin resonance, the enhancement is
completely removed by dynamical screening (see AW).

As derived in AW, the dynamical transverse susceptibility $\chi ^{+-}(\Omega
)$, where $\Omega $ is the frequency of an a.c.\ electromagnetic field
polarized transverse to the static magnetic field, is given by 
\begin{equation*}
\chi ^{+-}(\Omega )=\mu _{B}^{2}[g_{c}^{2}\chi _{cc}^{+-}(\Omega
)+g_{f}^{2}\chi _{ff}^{+-}(\Omega )+2g_{c}g_{f}\chi _{cf}^{+-}(\Omega )].
\end{equation*}
The partial susceptibilities are obtained by evaluating Feynman bubble
diagrams dressed by vertex corrections of the ladder type referring to the
Fermi liquid interaction (local electrons) and the spin-orbit interaction
(impurity correlation lines for the conduction electrons). The final result
obtained in AW may be reexpressed as 
\begin{widetext}
\begin{eqnarray*}
\chi ^{+-}(\Omega +i0) &=&\mu _{B}^{2}\{[(g_{f}\chi _{ff}^{+-}(0)+g_{c}\chi
_{cf}^{+-}(0))^{2}/\chi _{ff}^{+-}(0)]\frac{-\omega _{r}+i\Gamma}{\Omega
-\omega _{r}+i\Gamma} \\
&+& g_{c}^{2}\frac{-\omega _{\mathbf{k}_{F}}^{-}+2i\gamma _{\mathbf{k}%
_{F}}^{-}}{\Omega -\omega _{\mathbf{k}_{F}}^{-}+2i\gamma _{\mathbf{k}%
_{F}}^{-}}[\chi _{cc,H}^{+-}(0)-(\chi _{cf,H}^{+-}(0))^{2}/\chi
_{ff,H}^{+-}(0)]\}
\end{eqnarray*}%
\end{widetext}

There appear to be two different resonance denominators in the above
expression. Only the resonance at $\omega _{r}$ has been observed in ESR
experiments \ The resonance at $\omega _{\mathbf{k}_{F}}^{-}$, is shifted to
much higher frequencies (the factor $R$ !). However, a closer look reveals
that the weight of the resonance at $\omega _{\mathbf{k}_{F}}^{-}$ is zero
in the regime considered. Indeed, by inserting the quasiparticle weight
factors, $\chi _{ab,H}^{+-}(0)\simeq 2N_{0}(a_{\mathbf{k}_{F}\sigma
}^{ab,-})^{2}$ , where $a,b$ is $c\ $or $f$, one finds that the prefactor of
the second resonance is zero. In the limit of equal g-factors, the prefactor
of the first resonance simplifies to $2N_{0}R$ and its position is
unshifted, $\omega _{r}=\omega _{f}$ . In the absence of spin lattice
relaxation, the linewidth shrinks to zero in this case. This is not
reproduced in the above calculation, since we did not take into account the
vertex corrections belonging to the imaginary part of the self energy (the
scattering-in term in Boltzmann equation language).

\subsection{Non-spin rotation invariant Fermi liquid interaction}

The relatively large $g$-shift observed in experiment suggests the presence
of a small additional spin-symmetry breaking Fermi liquid interaction $%
\mathbf{I}$. Since this interaction can only be mediated by the spin-orbit
interaction it should have preferred direction given by the lattice
symmetry. We therefore assume the form \ $\mathbf{I}_{\alpha\beta
;\gamma\delta}=-I(\mathbf{c\cdot \boldsymbol{\tau } }_{\alpha\beta })(%
\mathbf{c\cdot \boldsymbol{\tau } }_{\gamma\delta})$, where $\boldsymbol{%
\tau }$ is the vector of Pauli matrices and $\mathbf{c}$ is a unit vector in
the direction of the crystallographic $c$-axis of the tetragonal lattice.
The screened and unscreened tensor susceptibilities $\mathbf{X},\mathbf{X}%
_{H}$, where $\mathbf{X} _{ij}=\langle\langle S_{i};S_{j}\rangle\rangle$, \
\ $i=$ $x,y,z$, are connected by the Bethe-Salpeter equation

\begin{equation*}
\mathbf{X} = \mathbf{X}_{H}+\widetilde{U}(\mathbf{X}_{H} \mathbf{X})+4I(%
\mathbf{X}_{H}\cdot \mathbf{c})(\mathbf{c\cdot }\mathbf{X}).
\end{equation*}%
The solution is given by

\begin{equation*}
\mathbf{X} =(1-\widetilde{U} \mathbf{X}_{p})^{-1} \mathbf{X}_{p}
\end{equation*}%
where $\mathbf{X}_{p}$ is the projected unscreened susceptibility

\begin{equation*}
\mathbf{X}_{p}=\mathbf{X}_{H}+\frac{4I}{1-4I\chi _{H}^{cc}}(\mathbf{X}%
_{H}\cdot \mathbf{c})(\mathbf{c\cdot }\mathbf{X}_{H}).,\text{ \ \ \ \ \ \ \
\ }\chi _{H}^{cc}=(\mathbf{c\cdot }\mathbf{X}_{H}\cdot \mathbf{c})\text{\ \
\ }
\end{equation*}%
To linear order in $I$ the general expression simplifies to 
\begin{equation*}
\mathbf{X}=[1-\widetilde{U}\mathbf{X}_{H}-4I\,\mathbf{X}_{H}^{-1}(\mathbf{X}%
_{H}\cdot \mathbf{c})(\mathbf{c\cdot }\mathbf{X}_{H})]^{-1}\mathbf{X}_{H}
\end{equation*}%
In the main configuration of the experiments, the static magnetic field is
oriented parallel to the $ab$-plane, say along the $a$-axis. We take this to
be the $z$-axis in spin space and identify the $c$-axis with the $x$-axis.
Then we see that the screening of the static field is not changed to linear
order in $I,$ as $\mathbf{X}_{+z}=\mathbf{X}_{z+}=0,$ etc. The dynamical
response with the time-dependent magnetic field oriented perpendicular to
the $z$-axis is, however modified. Using $\chi ^{+x}\chi ^{x-}=\frac{1}{4}%
(\chi ^{+-})^{2}$ we find

\begin{equation*}
\chi ^{+-}=\chi _{H}^{+-}/[1-(\widetilde{U}+I)\chi _{H}^{+-}]
\end{equation*}


\begin{thebibliography}{99}
\bibitem{expl} The term proportional to $m/m^*$ was inadvertently
misevaluated in Eq.\ (20) of AW.
\end{thebibliography}
\end{document}